\documentclass[reprint,amsmath,amssymb,aps,pra]{revtex4-2}

\usepackage{graphicx}
\usepackage{dcolumn}
\usepackage{bm}
\usepackage{color}
\usepackage{braket}  
\usepackage{subfigure}

\begin{document}

\title{Topological-Vacuum-Induced Strong Photon-Exciton  Coupling}%
\author{Yali Jia $^{1}$}
\author{Zihan Mo$^{1}$}
\author{Qi Liu$^{1,2}$}
\author{Yu Tian$^{1,2}$}
\author{Zhaohua Tian$^{1}$}
\author{Qihuang Gong$^{1,2,3,4,5}$}
\author{Ying Gu$^{1,2,3,4,5}$}
 \email{ygu@pku.edu.cn}

\affiliation{$^1$State Key Laboratory for Mesoscopic Physics, Department of Physics, Peking University, Beijing 100871, China\\
$^2$Frontiers Science Center for Nano-optoelectronics $\&$  Collaborative Innovation Center of Quantum Matter $\&$ Beijing Academy of Quantum Information Sciences, Peking University, Beijing 100871, China\\
$^3$Collaborative Innovation Center of Extreme Optics, Shanxi University, Taiyuan, Shanxi 030006, China\\
$^4$Peking University Yangtze Delta Institute of Optoelectronics, Nantong 226010, China\\
$^5$Hefei National Laboratory, Hefei 230088, China}

\date{\today}
\begin{abstract}
The electromagnetic vacuum construction based on micro-nano photonic structures is able  to engineer the photon-exciton interaction at the single quantum level. 
Here,  through engineering the electromagnetic vacuum background formed by edge states, we demonstrate a strong photon-exciton coupling in topological photonic crystal containing a dielectric nanoantenna.
By guiding the scattering photons into the edge states, the linewidth of nanoantenna with more than hundred nanometers in air can be reduced into only several nanometers due to topological robustness, 
so that both strong coupling condition and high photon collection efficiency can be achieved.
Electromagnetic vacuum background under topological protection holds great promise for controlling the light-matter interaction in quantum optics and  on-chip quantum information.

\begin{description}
\item[Keywords] cavity quantum electrodynamics; topological photonics; electromagnetic vacuum
\end{description}

\end{abstract}

\maketitle


Electromagnetic vacuum construction, including engineering the electromagnetic mode inside the cavity and the electromagnetic environment outside the cavity, plays an important role in the strong photon-exciton coupling at the single-quantum level.
The condition of strong coupling  is $g>\kappa,\gamma $, where $g$ is the coupling strength between emitter and the cavity mode, $\kappa$ the cavity loss, and $\gamma$ the decay rate of the emitter~\cite{vahala2003optical,li2016transformation}.
To achieve the strong coupling, researchers pay more attention  to  increase the coupling strength $g$ between cavity mode and exciton, which is determined by the optical mode volume~\cite{tame2013quantum,jacob2011plasmonics,savasta2010nanopolaritons}. 
While the electromagnetic vacuum background (TVB)  in which the cavity is located can effectively influence coupling strength $g$, the cavity loss $\kappa$ and decay rate $\gamma$~\cite{ren2017evanescent,peng2017enhancing,liberal2017zero,li2023highly}.
Owing to possessing diverse optical modes, micro-nano photonic structures play an indispensable role in electromagnetic vacuum construction~\cite{schlather2013near,sun2022empowering,zhang2019chiral}. 
Especially in their hybrid structures, one mode can act as  the cavity mode while the others  can be  regarded as electromagnetic vacuum environment~\cite{doeleman2016antenna,yan2015directional,liu2017nanoantenna,mcpolin2018imaging}.

In previous studies of strong coupling, optical modes in micro-nano photonic structures  are generally acted as the cavity modes, such as surface plasmons in metallic nanoparticles~\cite{chikkaraddy2016single,liu2017strong}, defect modes in photonic crystals (PC)~\cite{hennessy2007quantum,englund2007controlling,yoshie2004vacuum}, Mie resonances in dielectric nanoparticles~\cite{castellanos2020exciton,tserkezis2018mie}.
While in other studies, optical modes   can serve as an electromagnetic vacuum environment where the cavity is located, that is, through engineering  these modes outside the cavity, $g$, $\kappa$  and $\gamma$ can be greatly modified~\cite{ren2017evanescent,peng2017enhancing,li2023highly}.
The topological photonic structure possessing edge states with the properties of robustness and anti-scattering,  is a good candidate to realize the photon-exciton interaction~\cite{haldane2008possible,lu2014topological,mittal2018topological,blanco2018topological,he2019silicon}.
However, when the topological edge state acts as the cavity mode, owing to its large mode volume, the coupling strength $g$ is hardly larger than the cavity loss $\kappa$, making it difficult to reach the strong coupling regime.
In the following, we proposed that while these edge states are looked as electromagnetic vacuum environment in which the nanocavity locates, the strong photon-exciton coupling in the nanocavity can be achieved.

In this Letter, by putting a resonant dielectric nanoantenna into the  topological PC [Fig. 1(a)], a strong photon-exciton coupling induced by the TVB is theoretically demonstrated.
Here a gap dielectric nanoantenna acts as the cavity with nanoscale mode volume, while the topological edge states become an electromagnetic vacuum background.
For a dielectric nanoantenna locating in the air, i.e., plane waves are the electromagnetic background, because of the linewidth of hundreds of nanometers, the  strong photon-exciton coupling cannot be realized.
While the dielectric nanoantenna is inside the TVB, by guiding scattering photons into the edge state channel,  the linewidth of nanoantenna is reduced to only several nanometers, thus strong coupling condition $g>\kappa,\gamma $ is easily achieved. 
 In addition, more than $60\%$ of photons  can be collected by the edge states.
 The topological electromagnetic vacuum environment  proposed here  holds great promise for controlling the light-matter interaction under topological protection in quantum optics and on-chip quantum information.

\begin{figure*}[htbp]
\includegraphics[width=1\textwidth]{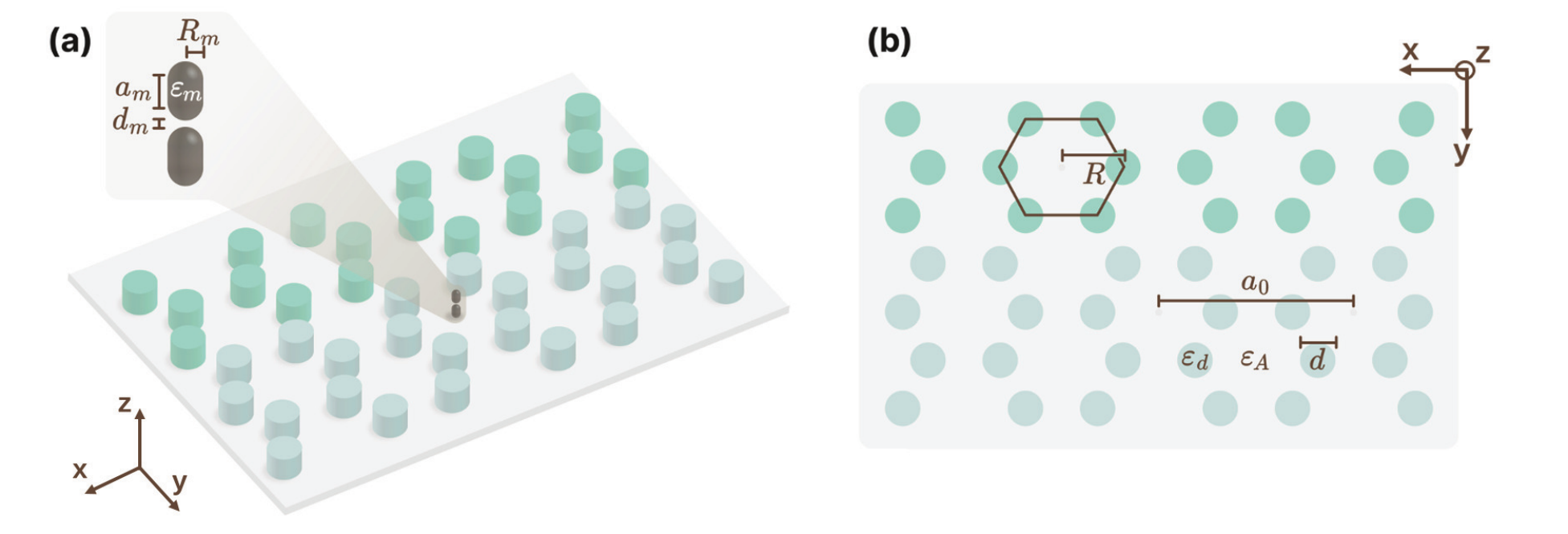}
\caption{ Schematic diagrams of (a)  dielectric  nanoantenna locating in the topological electromagnetic vacuum background and (b) topological photonic structure composed of two honeycomb PCs.
}\label{fig1}
\end{figure*}


The mechanism of narrowing linewidth of nanoatenna through TVB is described as follows.
For a dielectric nanoantenna locating in free space, the local electromagnetic field at the gap makes it have small mode volume, so the coupling coefficient $g$ can reach several meV.  
Because the photons scatter in all directions [Fig. 2(a)], there is a large scattering loss of about $2\times 10^3$ meV  [inset in Fig. 2(d)], so $g$ is much smaller than $\kappa$, namely, the strong coupling condition cannot be satisfied. 
While the nanoantenna is within the topological structure, i.e., it is in TVB formed by edge states,  $g$ remains almost unchanged  \cite{SM}. 
But now most scattering photons can be guided into the edge states [Figs. 2(b,c)], resulting in an ultra-narrow linewidth of several meV.  Owing to topological robustness, the linewidth of nanoantenna is the same as that of edge state in topological PC [Fig. 2(d)].
Therefore, since the TVB greatly suppresses the linewidth of nanoantenna, strong photon-exciton coupling is achieved.
While, if we put a silver nanoantenna into a topological structure, the part of intrinsic loss (or absorption) cannot be modified by TVB.  Hence, in this situation, by keeping similar linewidth as that in the air,  it is difficult to realize the strong coupling  \cite{SM}.

\begin{figure*}[htbp]
\includegraphics[width=1\textwidth]{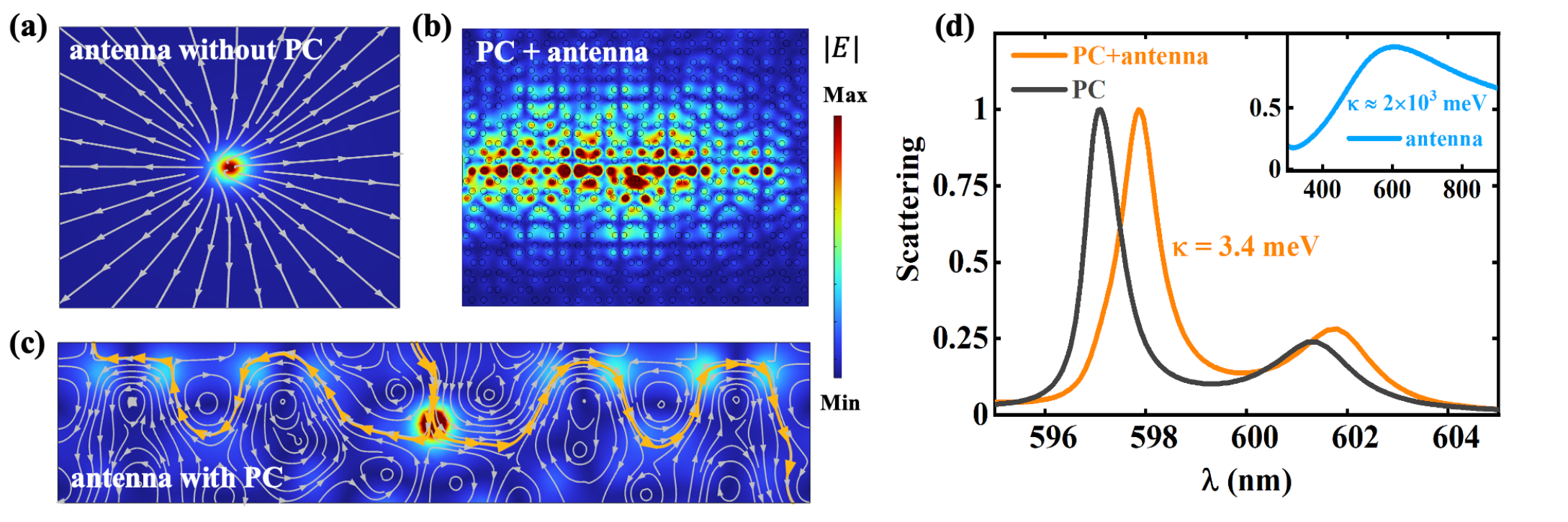}
\caption{ 
The mechanism of narrowing linewidths of a dielectric nanoantenna in TVB.
Electric field and energy flow distributions of  the nanoantenna  in the xy plane (a) without and (c) with topological PC.
(b) Electric field distribution of the edge state in hybrid topological  structure containing a nanoantenna. 
(d) The linewidths of edge state of topological PC (black curve) and nanoantenna in hybrid topological  structure (orange curve). The inset in (d) is the linewidth of nanoantenna in air.
}\label{fig2}
\end{figure*}

As shown in Fig. 1(a), the topological photonic structure is composed of two PCs with different topological properties. 
Topological edge states appearing at their interface have robustness to impurities and defects \cite{wu2015scheme}.
The edge state is a propagation mode with large optical mode volume. When it interacts with a single quantum emitter, the coupling coefficient $g$ is very small, which makes it difficult to reach the strong coupling regime. 
Here, we  regard these edge states as an electromagnetic vacuum environment, in which  a gap dielectric nanoantenna with a small mode volume is placed.
Therefore, by utilizing the large coupling coefficient brought by the strong local field at the nanoscale gap and the narrow linewidth of dielectric cavity in TVB, the strong photon-exciton  coupling can be achieved.


Such hybrid topological  structure is composed of two pieces of honeycomb PCs with $C_6$ symmetry and a dielectric nanoantenna.  
The two pieces of PCs have different topological properties, i.e., trivial and nontrivial.
As shown in Fig. 1(b), the brown hexagons are six cylinders forming a hexagonal honeycomb lattice unit cell, where $a_0$ is the lattice constant and $R$ is the side length of the hexagon. $d=65$ nm is the diameter of the cylinder,  $\varepsilon_d=11.7$ and  $\varepsilon_A=1$  are the dielectric constant of the cylinder and the surrounding environment, respectively.
Two kinds of PCs with different topological properties are $a_0/R = 2.7$  and $a_0/R = 3.3$.
With above parameters, the edge state appears at  about $\lambda=595\sim650$ nm.
Then we put the  dielectric nanoantenna into above topological structure, which is vertically located in the edge state channel. The parameters of nanoantenna are $R_m=45$ nm, $a_m=80$ nm, $d_m=5$ nm, and dielectric constant $\varepsilon_m=12$.
With these  parameters, its resonance wavelength  in the air is 610 nm.
We used COMSOL Multiphysics software to simulate the spectral properties of edge states of topological PC and the electric dipole resonance of nanoantennas \cite{SM}.


 From Fig. 2(d), it is seen that the linewidth of the nanoantenna in TVB is  same as that of the edge state in bare topological PC, which is only  $1$ nm, corresponding to the loss factor of $\kappa=3.4$ meV. 
Compared with the linewidth of about $2\times 10^3$ meV of the nanoantenna  in air [inset of Fig. 2(d)], it is narrowed several  hundred times.  
The reason is that edge state channels can collect photons those previously scattering in all directions.
Superior to the case of evanescent vacuum with low collection efficiency \cite{ren2017evanescent}, the high collection efficiency here  originates from the robustness of topological edge states. 
It is noted that the narrowing of nanoantenna linewidth in TVB is also the result of topological state-led mode coupling between edge states and dielectric resonance, as proposed in Ref. \cite{qian2021absorption}.


We now focus on the strong coupling parameters $g, \kappa, \gamma $  for a dielectric nanoantenna in the TVB.
As shown in Fig. 3(a), owing to strong local field at the nanoscale gap, the photon-exciton coupling coefficient $g$  can reach $5.46$ meV, which is almost the same as that of the antenna in air \cite{SM}. 
Here $g=\vec{E} \cdot \vec{\mu } /\hslash$, where $\mu=0.2$ enm, and $\vec{E} $ is the electric field corresponding to the single-photon excitation of the nanocavity in the TVB.
    Simultaneously, the loss coefficient of the antenna in the topological structure  is only $\kappa=3.4$ meV  [Fig. 2(d)], which is much smaller than the cavity loss $\kappa\approx 2 \times 10^3$ meV in air;
 the decay rate of the quantum emitter $\gamma=0.07\mu$eV when there are only topological edge states.
So the strong coupling condition $g>\kappa,\gamma$ is fully satisfied.
More calculation details are shown in Ref. \cite{SM}.

\begin{figure}[htbp]
\includegraphics[width=0.5\textwidth]{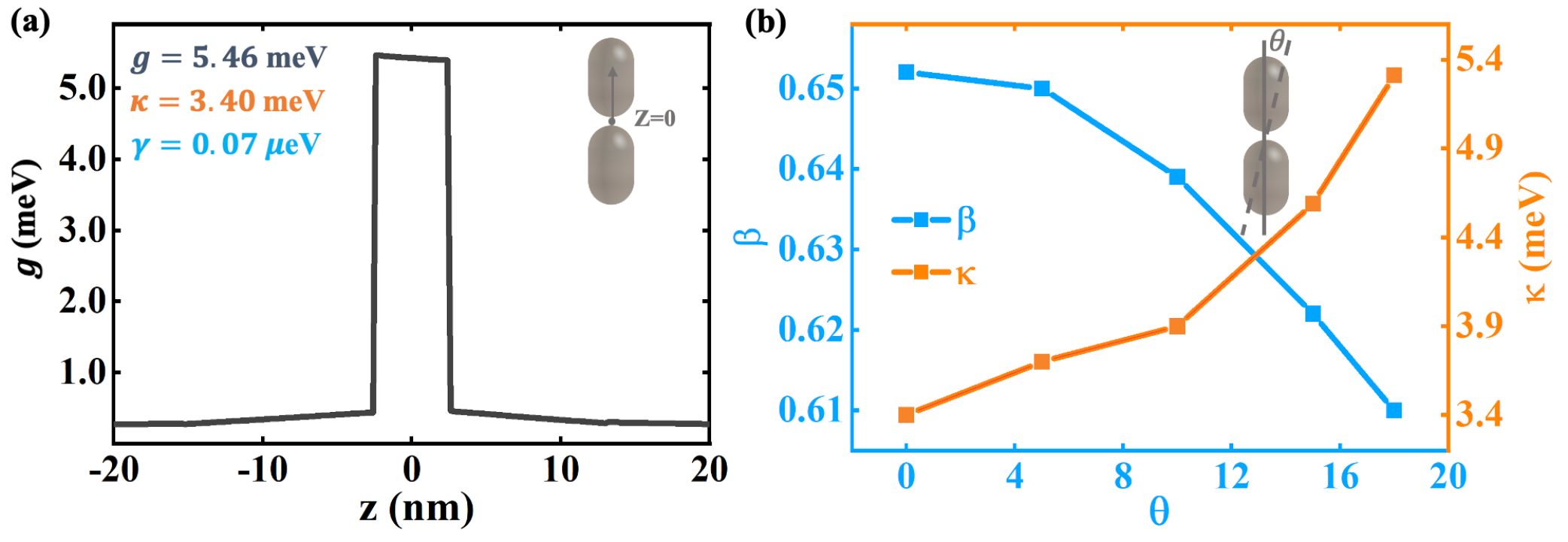}
\caption{(a) Coupling coefficient $g$ between the quantum emitter and the nanoantenna locating in the TVB. (b) Cavity loss $\kappa$ and photon collection efficiency $\beta$ varying with the angle $\theta$.
Here $d_m$=5 nm and $\theta$ is angle between the antenna and z-axis.
}\label{fig3}
\end{figure}

Next, we explore the fluorescence collection through topological edge states.
For a nanoantenna in free space, after the interaction between the photon and exciton, the resonance fluorescence scatters  in all directions, so it is difficult to be collected and utilized \cite{Regmi}.
While, if a nanoantenna is placed in topological photonic structure, the fluorescence can be guided into the edge state.
The collection efficiency by edge state is defined as $\beta=\gamma_{ed}/\gamma_{tot}$, where $\gamma_{ed}$ is the part that flows into the edge state, and $\gamma_{tot}$ is the total emission rate \cite{SM}.
Owing to the topological robustness, the collection efficiency $\beta$ can reach 65\%,  which is much higher than that of evanescent waves in nanowires~\cite{ren2017evanescent,R.Yalla2012}.
Photons collected and propagated through topological channels can be used for on-chip quantum device design and quantum information processes.

In real nanofabrication process, it is generally difficult to  vertically embed a nanoantenna into topological photonic structure, i.e., there will be a slight angle $\theta$ between the nanoantenna and the z-axis. 
So we need to discuss the variation of loss coefficient $\kappa$  and fluorescence collection efficiency $\beta$  with $\theta$.
 As shown in Fig. 3(b), as $\theta$  varies from $0^o$ to $18^o$, there is a decrease in $\beta$,  but an increase  in $\kappa$.  
It is found that even though the antenna placed into the topological PC is slightly tilted, the photon-exciton strong coupling  as well as the efficient fluorescence collection can still be remained. 



The effect on the strong coupling  and fluorescence collection is investigated when the structural parameters of the antenna is changed. 
It is found that when the radius  $R_m$ and  the gap distance $d_m$ vary slightly, $g, \kappa$  and $\gamma$  are almost unchanged, except that $g$ changed from 4 to 6 meV as a function of $d_m$ [Fig. 4], which means that strong coupling maintains very well.
The fluorescence collection efficiency $\beta$ is always kept at around 63\%, which is insensitive to the change of structural parameters \cite{SM}.
The above results illustrate the robustness of strong coupling properties in topological hybrid structure, which will be beneficial to the experimental implementation.

\begin{figure}[htbp]
\includegraphics[width=0.5\textwidth]{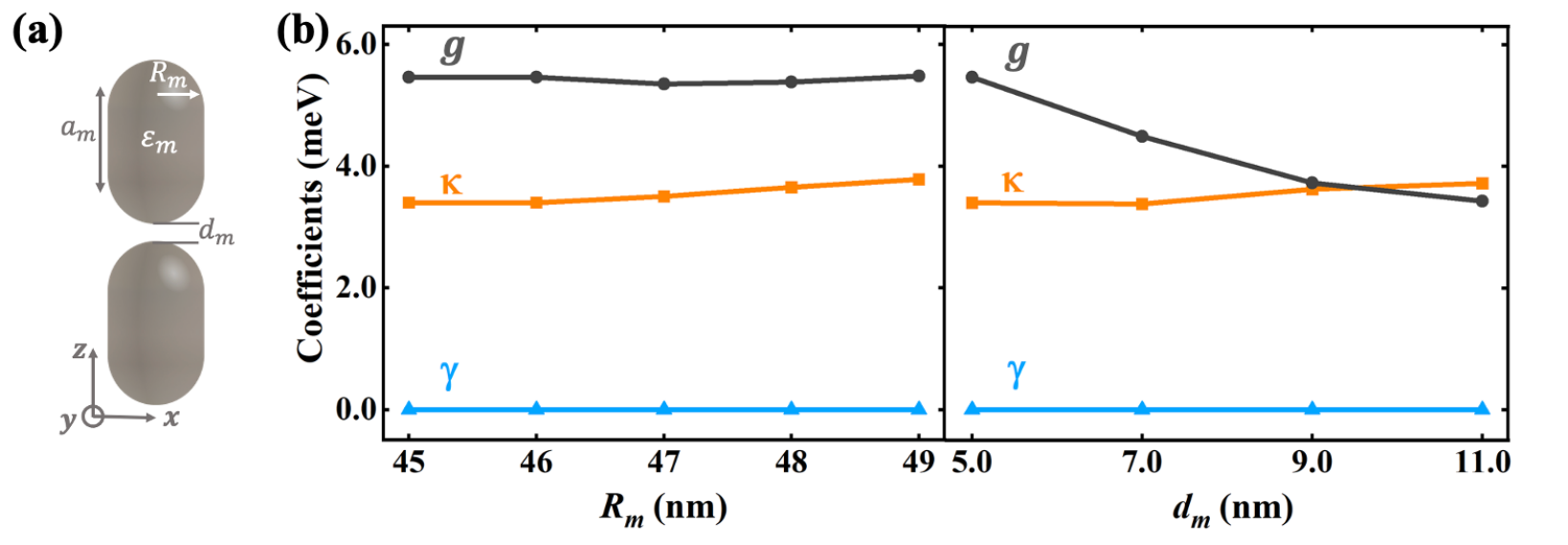}
\caption{ The coefficients $g,\kappa,\gamma$ as a function of $R_m$ and $d_m$ of the nanoantenna in TVB.}
\end{figure}

Finally, we discuss the possibility of experimental implementation of the scheme proposed here.
At present, both topological PC~\cite{T.O} and dielectric nanocavities ~\cite{D. G2017} can be manufactured using nanotechnology, e.g., electron beam lithography and electrochemical etching.
It is also possible to insert emitters in nanoscale gaps in our structure, for example the arrangement of terphenyl terrylene molecules has been achieved in spin-coated ultrathin crystal films of p-terphenyl~\cite{R. Pfab2004}.
The excited state of the emitter can be prepared by resonance of the nanoantenna in the edge state vacuum.
Edge states in topological PC can be excited via grating coupling~\cite{M. I.2019}.
Therefore, using existing techniques, it is possible to experimentally realize strong coupling in the dielectric cavity under topological vacuum.

In summary, by proposing the concept of topological electromagnetic vacuum, we have theoretically demonstrated the strong photon-exciton coupling in the topological structure containing a dielectric nanoantenna.
 Owing to the existence of TVB, the linewidth of nanoantenna is greatly suppressed, up to less one percent of the linewidth in air, so that the strong coupling condition is fully satisfied. 
 Also with the TVB,  the resonance fluorescence is collected  with high efficiency.
The concept of  topological vacuum proposed here can be  extended to other hybrid topological photonic structures, in which not only  the threshold for achieving strong coupling will be lowered, but also  a relatively stable environment can be provided.
Engineering the electromagnetic vacuum with topological robustness offers significant potential for nanoscale light-matter interactions manipulation and efficient on-chip quantum information processing.

\hspace*{\fill}\

This work is supported by the National Natural Science Foundation of China under Grants No. 11974032 and the Innovation Program for Quantum Science and Technology under Grant No. 2021ZD0301500.


\end{document}